\documentclass[a4paper,11pt]{article}
\usepackage{pos}

\title{Heavy-flavor hadronization mechanism from pp to AA collisions: a theoretical perspective}

\author*[a]{Andrea Beraudo}

\affiliation[a]{Istituto Nazionale di Fisica Nucleare - Sezione di Torino,\\
  Via Pietro Giuria 1, I-10125, Torino}

\emailAdd{beraudo@to.infn.it}

\abstract{The interest in studying heavy-flavor hadronization in high-energy nuclear collisions is twofold. On one hand hadronization represents a source of systematic uncertainties in phenomenological attempts of extracting heavy-flavor transport coefficients in the Quark Gluon Plasma which one assumes to be produced in the collision. Hence, developing the most possible reliable model for this process is important to get a precise and accurate estimate of a fundamental property of hot QCD. On the other hand studying how hadronization changes in the presence of a dense medium of colored partons can be considered an issue of interest by itself. In particular, the observation of modifications of heavy-flavor hadronization in proton-proton collisions strongly suggests that also in this case a small droplet of Quark-Gluon Plasma can be formed. Here we try to provide a general overview on heavy-flavor hadronization, from pp to AA collisions, stressing the aspects and challenges common to all mechanisms proposed in the literature. Then, focusing on a particular model, we show how a consistent description of several observables involving heavy-flavor hadrons can be obtained.}

\FullConference{
 26-31 March 2023 \\
 Aschaffenburg, Germany\\}

\begin{document}
\maketitle

\section{Motivations: why studying heavy-flavor hadronization?}
\begin{figure*}
\centering
\includegraphics[clip,height=5cm]{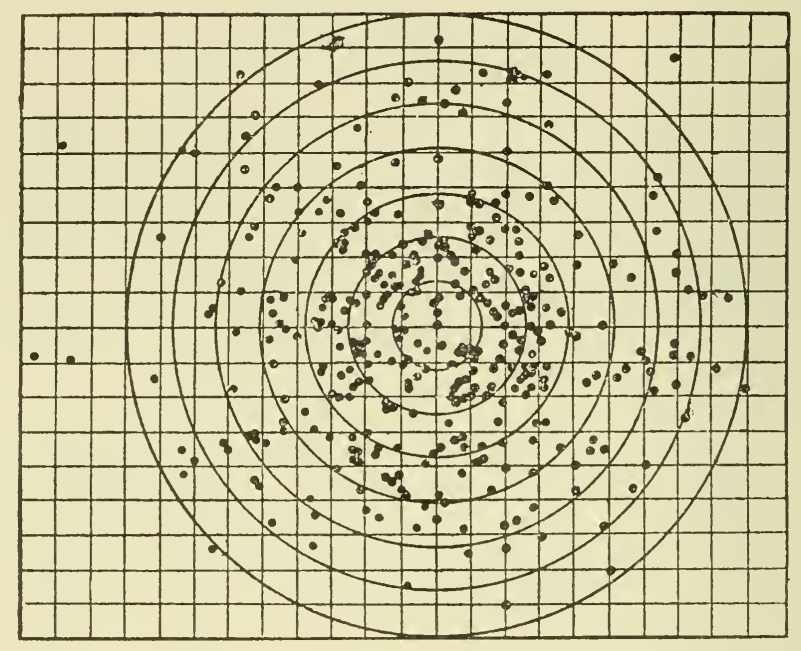}
\includegraphics[clip,height=5cm]{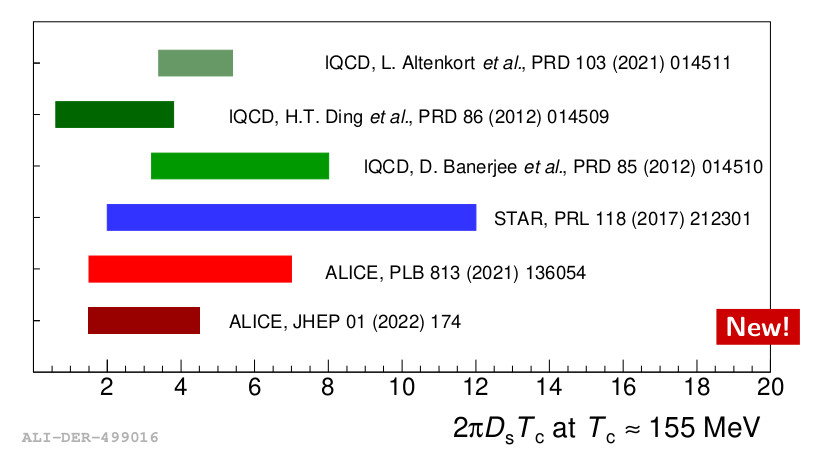}
\caption{Left panel: original drawing by Perrin studying the diffusion of colloidal particles in water~\cite{Perrin:1926}. Right panel: different estimates of the heavy-quark spatial diffusion coefficient in the QGP.}\label{fig:diffusion}
\end{figure*}
The original motivation for studying in-medium heavy-flavor (HF) hadronization was related to the extraction of the transport coefficients governing the heavy-quark dynamics in the hot, deconfined and expanding fireball produced in relativistic heavy-ion collisions (HIC's). Within this setup heavy quarks (HQ's) are described as Brownian particles undergoing a stochastic dynamics in the Quark-Gluon Plasma (QGP), modelled through some transport equation (Boltzmann, Fokker-Planck or Langevin), until they reach a hadronization hypersurface where they give rise to the final HF hadrons. Exploiting the stochastic diffusive dynamics of Brownian particles to access information on microscopic medium properties is not a novelty. More than 100 years ago the study of the diffusion of colloidal particles in water allowed Perrin to get a quite precise and accurate estimate for the Avogadro number, ${\cal N}_{\rm A}\!\approx\!5.5-7.2\cdot 10^{23}$~\cite{Perrin:1926}. Nowadays, in nuclear collisions, HF studies aim at quantifying with a similar precision and accuracy the HQ spatial ($D_s$) and momentum ($\kappa$) diffusion coefficients. Recent results are shown in Fig.~\ref{fig:diffusion}. As one can see, we are still quite far from achieving this goal. One of the major systematic uncertainties affecting phenomenological estimates is represented by hadronization, since the final detected particles are not the parent HQ's, but their daughter hadrons: hence the interest in developing the most possible reliable description of this process.

Indeed, understanding how the quark-to-hadron conversion changes moving from a dilute to a dense system, with a lot of color charges floating around, can represent an item of interest by itself. In this connection, the study of open HF has the great advantage that one knows exactly one of the parent partons of the final hadron, which can only be a charm or beauty quark produced in a hard scattering occurring before the fireball starts its hydrodynamic expansion. Furthermore, the fact that the same modifications of HF hadrochemistry -- with a relative enhancement of charmed-baryon production -- measured in nucleus-nucleus collisions are also observed in the proton-proton case~\cite{ALICE:2020wfu} may be considered a signature that also in these smaller systems a little droplet of QGP is produced.

\section{Common features and challenges to any hadronization model}\label{sec:common}
\begin{figure*}
\centering
\includegraphics[clip,height=5cm]{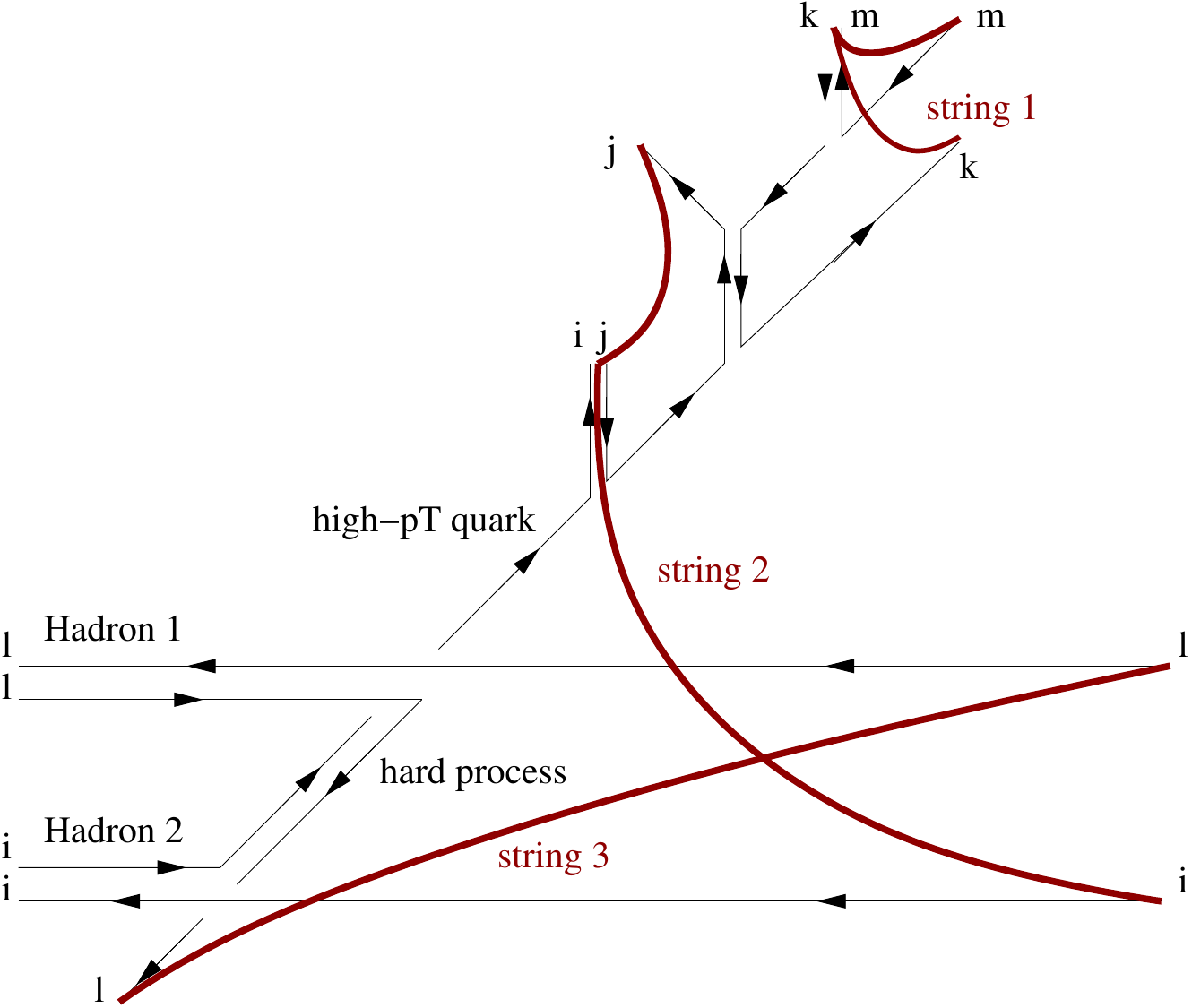}
\hspace{1.5cm}
\includegraphics[clip,height=5cm]{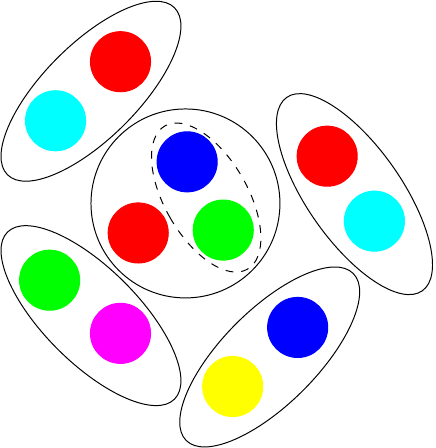}
\caption{Composite color-singlet objects from which hadronization starts in different models. Left panel: strings stretched following the color-flow in a hadronic collision. Right panel: clusters built recombining the closest opposite color charges (quarks or diquarks) in a dense partonic environment.}\label{fig:common}
\end{figure*}
Any conceivable hadronization mechanism must start from grouping colored partons into color-singlet structures. Depending on the model, these composite objects are referred to as strings (e.g. in PYTHIA), clusters (e.g. in HERWIG) or are directly identified with the final hadrons/resonances (as in coalescence models). What changes in the different physical situations is where the recombining partons are taken from. In collisions leading to the formation of a sufficiently dilute system (left panel of Fig.~\ref{fig:common}) the partons are taken from the hard scattering, from the parton-shower stage, from the underlying event and from the beam remnants and are grouped following the color-flow of the event. If, on the contrary, a very dense partonic system is formed -- as in HIC's -- one can assume that color neutralization occurs locally (right panel of Fig.~\ref{fig:common}), with the considered parton (the HQ in this case) undergoing recombination with the closest opposite color-charge. As suggested by the figure, the latter can also be a diquark, which favors the formation of clusters carrying one unit of baryon number. Notice that recombining nearby partons, besides minimizing the potential energy stored in the color-field, entails a strong space-momentum correlation (SMC), since particles belonging to the same fluid cell of an expanding fireball must share a common collective velocity. This, as we will see, has deep consequences for the typical invariant-mass of the formed clusters and for the kinematic distributions of the final hadrons. 

\begin{figure*}
\centering
\includegraphics[clip,height=3.5cm]{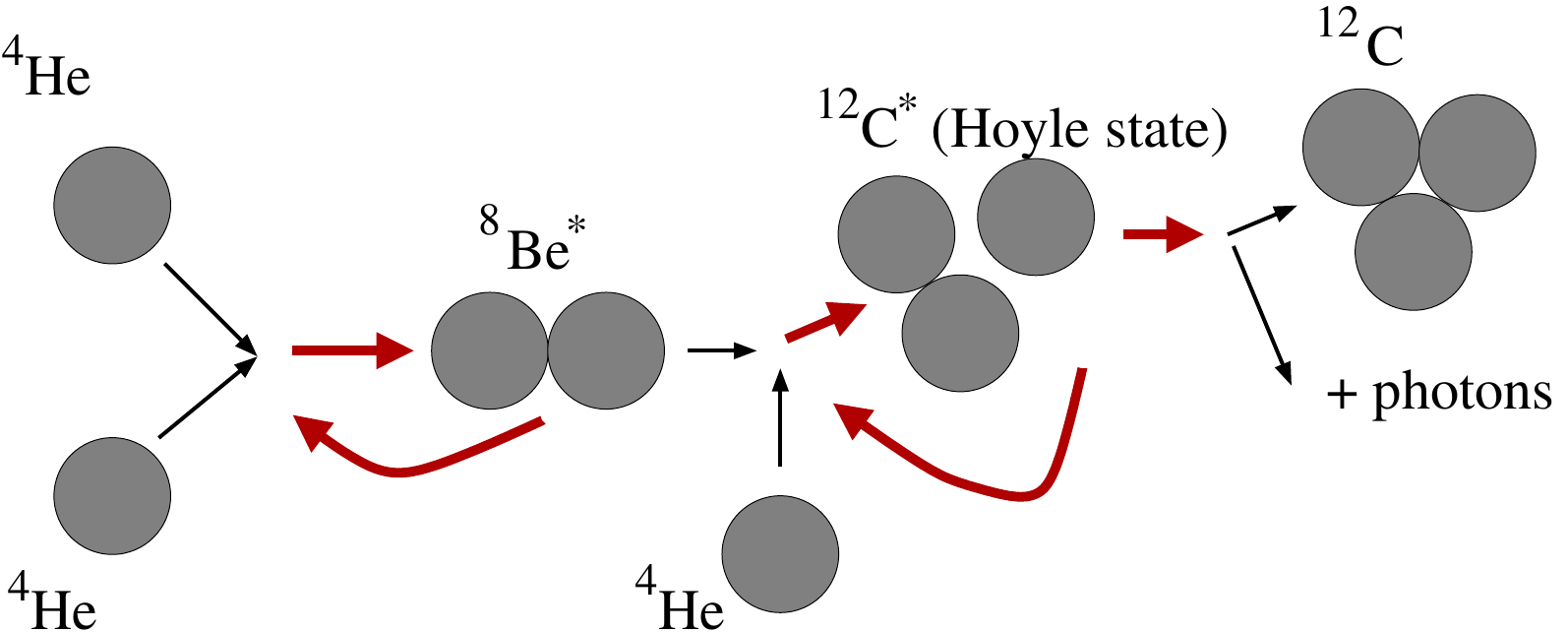}
\includegraphics[clip,height=3.5cm]{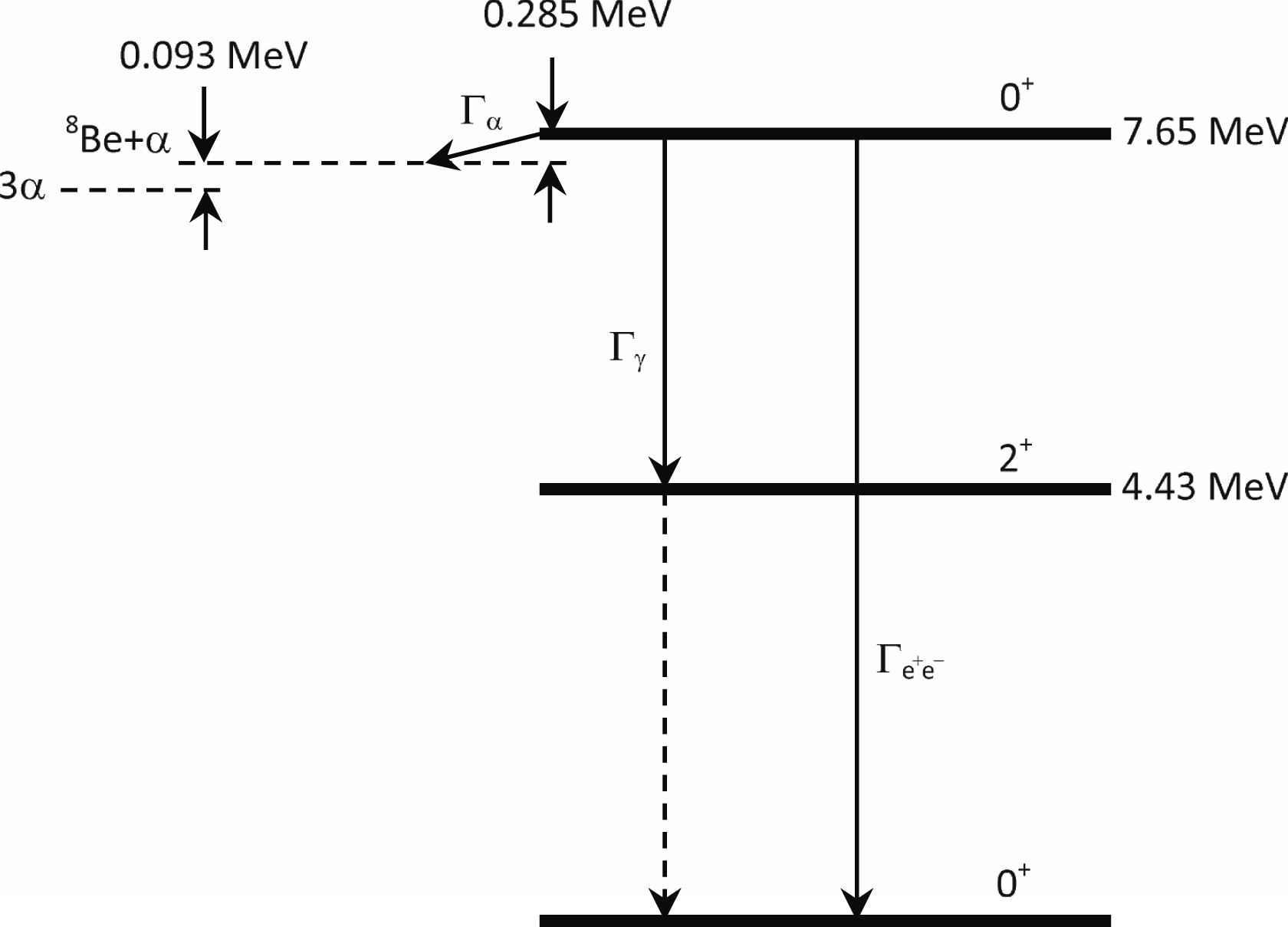}
\caption{The steps involved in the stellar nucleosynthesis of $^{12}$C, whose energy spectrum with an excited state just above the $^8$Be*+$^4$He threshold favors their recombination.}\label{fig:stellar}
\end{figure*}
In order to appreciate the challenges in developing a realistic model of hadronization it may be useful to consider a different situation in which composite objects are formed starting from more elementary degrees of freedom: stellar nucleosynthesis of $^{12}$C (see Fig.~\ref{fig:stellar}). $^{12}$C can be considered a cluster of 3 $\alpha$ particles, taken as the elementary building blocks of the process. They are the equivalent of quarks in hadronization. A direct recombination of 3 $\alpha$ particles would be very unlikely, but the process is favored by the existence of a resonant $^8$Be* state just above the 2$\alpha$ threshold, produced in the reaction $\alpha+\alpha\leftrightarrow ^8$Be* and which can live long enough, until the scattering with a third $\alpha$ particle. $^8$Be* can be considered the equivalent of diquarks in hadronization, which favor the formation of three-quark clusters. However, also in this case, the direct formation of $^{12}$C would be extremely unlikely. Thus, the great abundance of $^{12}$C in the Universe led Hoyle to predict the existence of an excited $^{12}$C* state just above the  $\alpha+^8$Be* threshold, easily accessible in a scattering process~\cite{Hoyle:1954zz}. In a tiny fraction of cases  $^{12}$C* undergoes an electromagnetic decay into the ground-state, explaining the observed carbon abundance. Soon after its prediction $^{12}$C* was actually discovered~\cite{PhysRev.92.649}. Clearly the latter can be considered the equivalent of the long list of excited hadronic resonances predicted by relativistic quark models (RQM's), whose feed-down is necessary to explain the observed abundance of ground-state HF hadrons in high-energy collisions within calculations assuming their statistical production from a hadronization hypersurface~\cite{He:2019tik}.

\begin{figure*}
\centering
\includegraphics[clip,height=4cm]{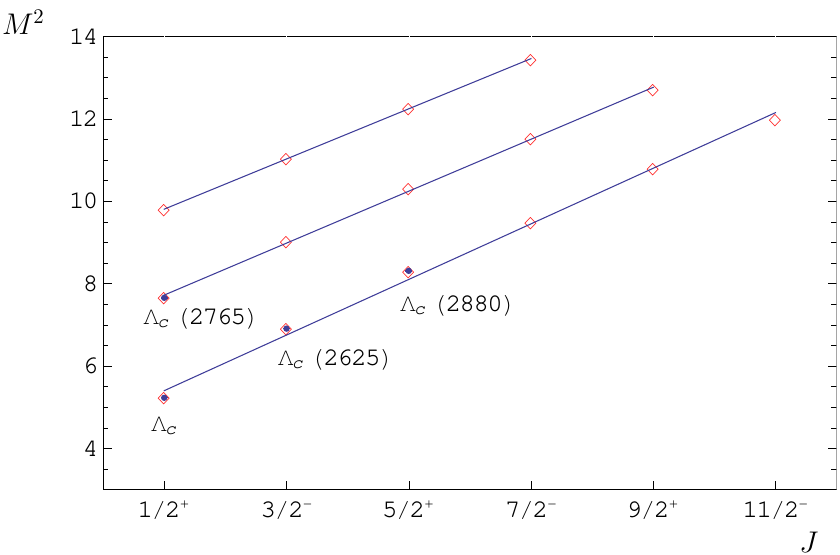}
\includegraphics[clip,height=4cm]{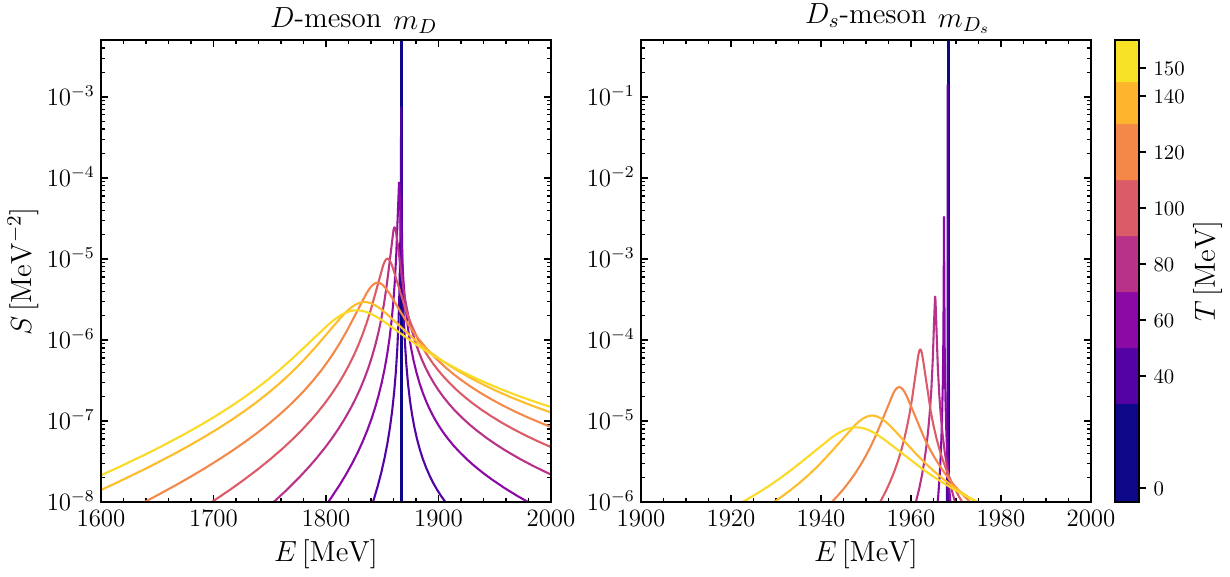}
\caption{Left panel: Regge trajectories for charmed baryons provided by the RQM of Ref.~\cite{Ebert:2011kk}. Only a fraction of the predicted states has been so far observed and listed in the PDG tables. Right panel: in-medium D-meson spectral functions provided by the EFT Lagrangian of Ref.~\cite{Montana:2020lfi}.}\label{fig:spectral}
\end{figure*}
To summarize, the final yields of stable nuclei in stellar nucleosynthesis are extremely sensitive to the existence of excited states just above the two-particle threshold, which have been experimentally well know for long and which are also predicted by theory calculations~\cite{Epelbaum:2011md}. Furthermore, the stellar temperatures $\sim\! 10^8 K\!\sim\! 10$ keV are not high enough to affect the nucleon/nuclear properties. Unfortunately, none of the above conditions is actually satisfied in the quark-to-hadron transition in HIC's. First of all, only few of the hadronic resonances predicted by the RQM ~\cite{Ebert:2011kk} and necessary to reproduce the measured yields of ground-state HF hadrons have been so far experimentally observed and listed in the PDG tables (left panel of Fig.~\ref{fig:spectral}). Secondly, at temperatures around the QCD crossover hadron spectral functions are strongly modified, both in the light~\cite{Hatsuda:1985eb} and in the heavy~\cite{Montana:2020lfi} sectors (right panel of Fig.~\ref{fig:spectral}), raising questions about the nature itself of a hadron around $T_c$. A bound state in the vacuum can remain a bound state, but can also become a broad resonance above the two-particle threshold. In this connection, the Resonance Recombination Model (RRM) developed in Ref.~\cite{Ravagli:2007xx} precisely relies on the existence of resonant mesonic states $M$ above the $m_Q+m_{\overline q}$ threshold, whose large thermal width allows the reactions $Q\overline q\leftrightarrow M$ to reach dynamical equilibrium.

\section{Recombination of color-charges in the QGP}
\begin{figure*}
\centering
\includegraphics[clip,height=4.5cm]{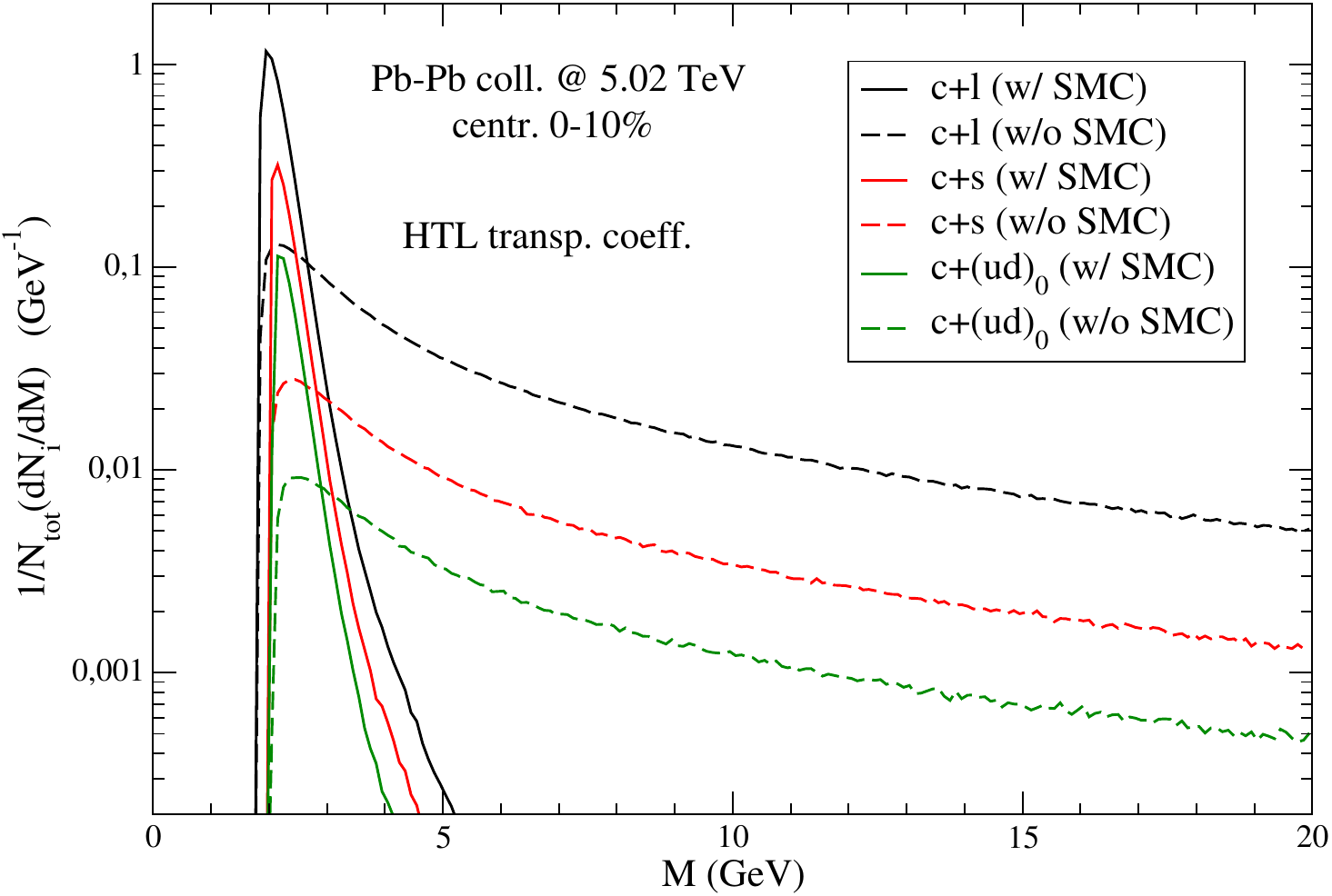}
\hspace{0.3cm}
\includegraphics[clip,height=4.5cm]{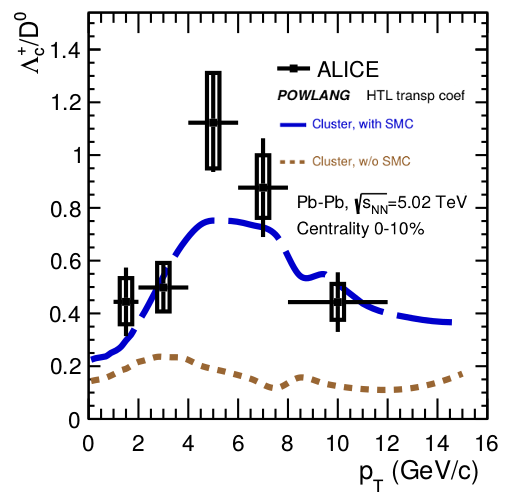}
\caption{The role of SMC in HF hadronization in HIC's. Left panel: invariant-mass distribution of various charmed color-singlet clusters. Right panel: $\Lambda_c^+/D^0$ ratio in Pb-Pb collisions.}\label{fig:SMC}
\end{figure*}
Since, as above mentioned, the modelling of the quark-to-hadron transition is so challenging, it is useful to keep the discussion as simple as possible, considering in deeper detail the minimal model of hadronization briefly sketched in Sec.~\ref{sec:common}. This will be sufficient to illustrate very general features of models based on parton recombination and to provide a quantitative guidance to interpret several experimental measurements. Within such a model, presented in Ref.~\cite{Beraudo:2022dpz}, color-neutralization occurs locally, on a isothermal hypersurface around the critical temperature $T_H=155$ MeV, recombining a HQ with an opposite color-charge from the same fluid cell, either a light antiquark of a diquark. Both the species and the momentum of the heavy-quark companion are sampled from a thermal distribution in the local rest frame of the fluid. A color-singlet cluster is then constructed, which tipically has quite a low invariant-mass, due to SMC which favors the recombination of collinear partons following the collective expansion of the firaball (see left panel of Fig.~\ref{fig:SMC}). Light clusters undergo then a 2-body decay into a charmed hadron plus a soft particle (tipically a pion), ensuring exact four-momentum conservation (at variance with standard coalescence approaches). Heavier clusters, above an invariant mass around 4 GeV, are treated as Lund strings whose fragmentation is simulated through PYTHIA 6.4. The recombination with diquarks -- assumed to be present in the fireball around the QCD transition -- enhances the production of HF baryon as compared to standard vacuum fragmentation (see right panel of Fig.~\ref{fig:SMC}). Notice that breaking SMC leads to a harder cluster-mass distribution and hence to a drop in the $\Lambda_c^+/D^0$ ratio, since high invariant-mass strings fragment as in the vacuum. Further results are discussed in the next Section, where a unified description of in-medium hadronization from pp to AA collisions is provided.

\section{In-medium hadronization also in small systems?}
\begin{figure*}
\centering
\includegraphics[clip,width=0.46\textwidth]{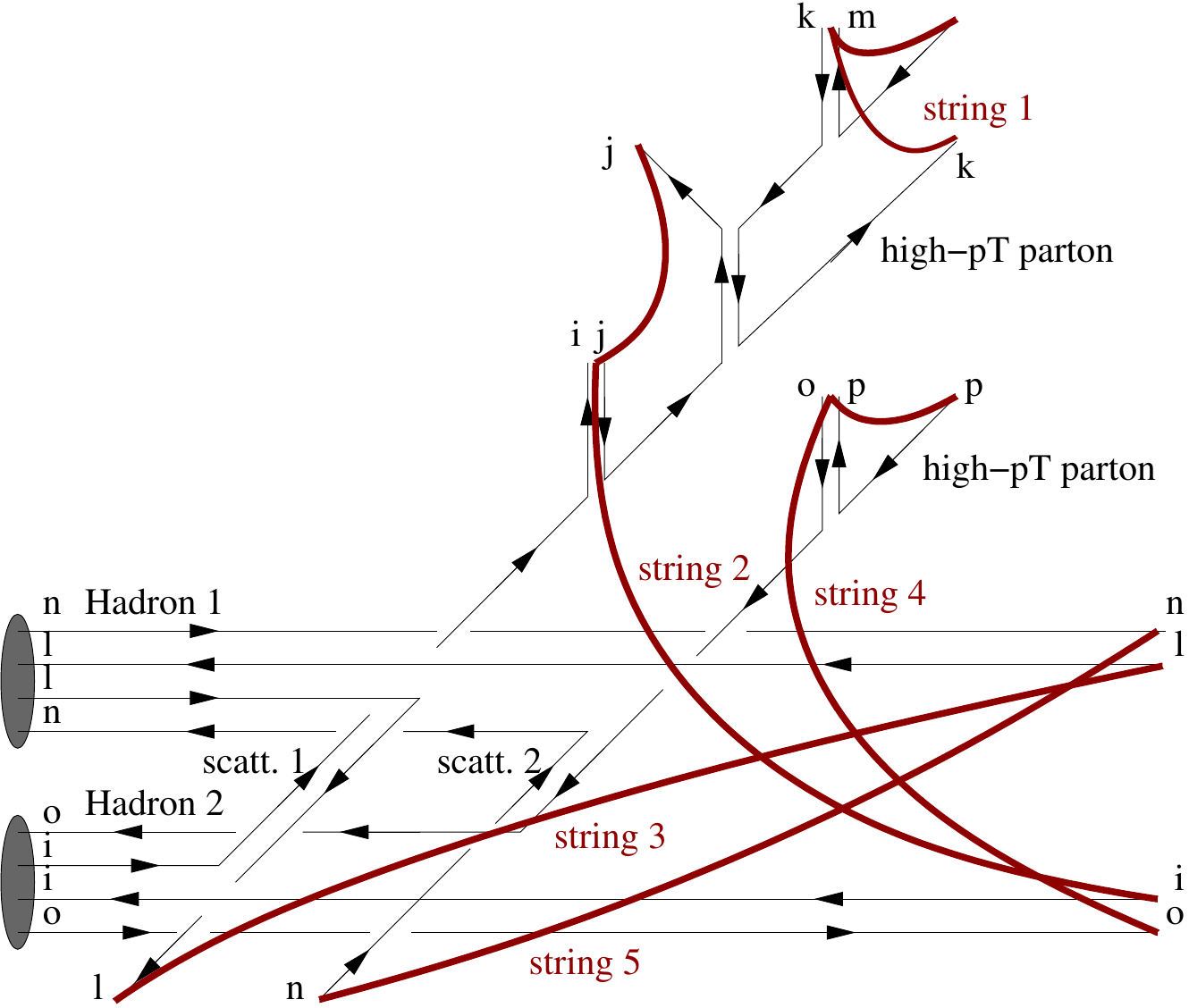}
\hspace{0.3cm}
\includegraphics[clip,width=0.46\textwidth]{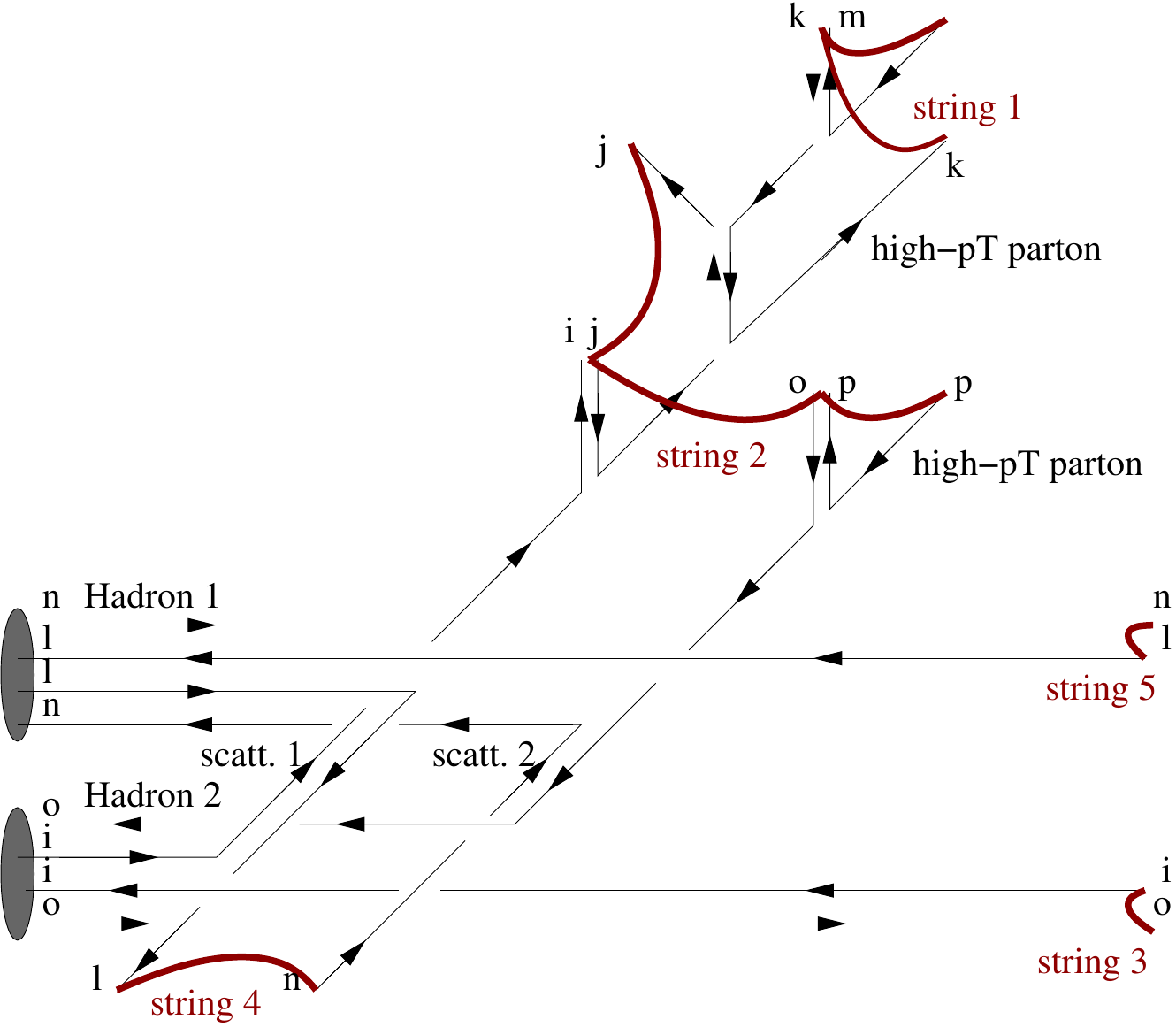}
\caption{Strings stretched between outgoing partons in a schematic hadronic collision before (left panel) and after (right panel) Color-Reconnection.}\label{Fig:CR}
\end{figure*}
One of the most surprising results involving HF observables obtained in proton-proton collisions is large value of the $\Lambda_c^+/D^0$ ratio~\cite{ALICE:2020wfu}, strongly enhanced with respect to expectations based on fragmentation fractions extracted from $e^+e^-$ data and compatible with measurements obtained in heavy-ion collisions. 
{Since in this last case the charmed-baryon enhancement is commonly attributed to a recombination process between the HQ and an opposite color-charge (possibly a diquark) from the hot, deconfined medium generated after the collision, one wonders whether a similar mechanism of hadronization can occur in proton-proton collisions, in which a small droplet of Quark-Gluon-Plasma (QGP) might also be produced.}
This was the idea proposed for instance in Refs.~\cite{Minissale:2020bif,Song:2018tpv}, which allowed the authors to satisfactory describe the $\Lambda_c^+/D^0$ ratio measured in pp collisions at the LHC. Another attempt to interpret the enhanced production of charmed baryons was based on the Statistical Hadronization Model~\cite{He:2019tik}, assuming a thermal population of the different charmed meson and baryon states predicted by the Relativistic Quark Model around a universal hadronization temperature, as already discussed in Sec.~\ref{sec:common}.
Reproducing such observations is a challenge for QCD event generators, but recent Color-Reconnection (CR) models implemented in PYTHIA 8~\cite{Christiansen:2015yqa} can provide a satisfactory description of the data. The mechanism is illustrated in Fig.~\ref{Fig:CR}. Strings, i.e. color flux-tubes, have a non-zero transverse thickness, with a radius around 0.5 fm~\cite{Baker:2018mhw}. Hence, in a hadronic collision with multiple partonic interactions, some of the strings stretched between the produced partons and the beam remnants may overlap and/or interact, leading to a rearrangement of the confining potential among the partons before hadronization which decreases the energy stored in the color field and favors the production of baryons. Notice that, even if strictly speaking this CR mechanism does not involve the formation of a deconfined medium, what occurs at hadronization is very similar. In fact, the color reconnections which tend to occur are the ones leading to a decrease of the invariant mass of the strings, hence the ones resulting in quite collinear partons as final string endpoints. But this is exactly what occurs, within a hot expanding fireball, in parton recombination approaches implementing SMC: one can consider them an extreme case of CR, in which the memory of the initial color connections is completely lost.

\begin{figure*}
\centering
\includegraphics[clip,height=4.5cm]{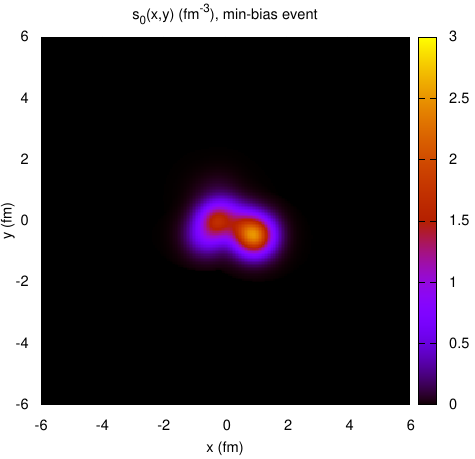}
\hspace{0.3cm}
\includegraphics[clip,height=4.5cm]{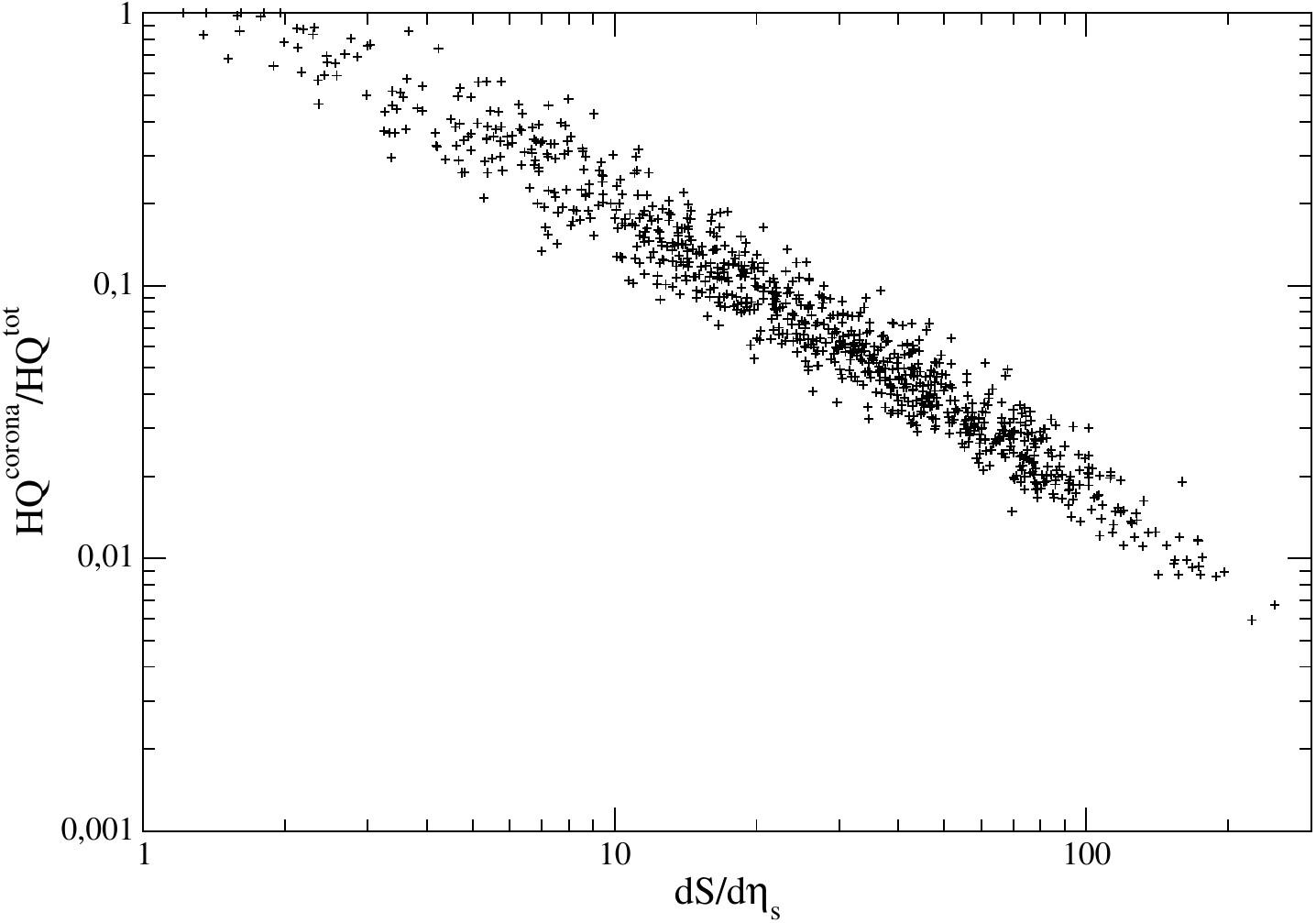}
\caption{Left panel: initial condition for a pp event from the minimum-bias sample. Right panel: fraction of HQ's initially found in fluid cells below $T_H$ in the various pp events of the minimum-bias sample, characterized by a different amount of deposited entropy per unit rapidity.}\label{Fig:pp-initial}
\end{figure*}
How reasonable is the assumption that the modification of HF production in proton-proton collisions depends on the formation of a small droplet of QGP in which HQ's undergo rescattering and recombination? This issue was addressed in detail in Ref.~\cite{Beraudo:2023nlq}, where the authors performed event-by-event (EBE) simulations involving the generation of a sample of minimum-bias initial conditions arising from the pp collisions, their hydrodynamic evolution, the stochastic propagation of HQ's throughout the fireball and their recombination with light thermal quarks or diquarks once reaching a hadronization hypersurface at $T_H\!=\!155$ MeV.
The initial entropy deposition in the transverse plane was modelled through the TrENTo code~\cite{Moreland:2014oya} and the authors checked that, on an EBE basis, one correctly reproduces the measured charged-particle multiplicity per unit rapidity. An example of initial condition from the minimum-bias sample is displayed in the left panel of Fig.~\ref{Fig:pp-initial}. The HQ's, generated with the the POWHEG-BOX package~\cite{Alioli:2010xd}, were then distributed among the different pp events according to the initial entropy-density in the transverse plane. Hence, they tend to populate the hot spots of the events with the largest deposited entropy per unit rapidity $dS/d\eta_s$. As a result, considering the minimum-bias sample, only a small fraction around 5\% of the HQ's is initially found in the fireball corona below $T_H$ (see right panel of Fig.~\ref{Fig:pp-initial}). Thus, it not surprising that the same modifications of HF production found in HIC's and attributed to a hot deconfined medium are also observed in the proton-proton case. 

\begin{figure*}
\centering
\includegraphics[clip,width=0.95\textwidth]{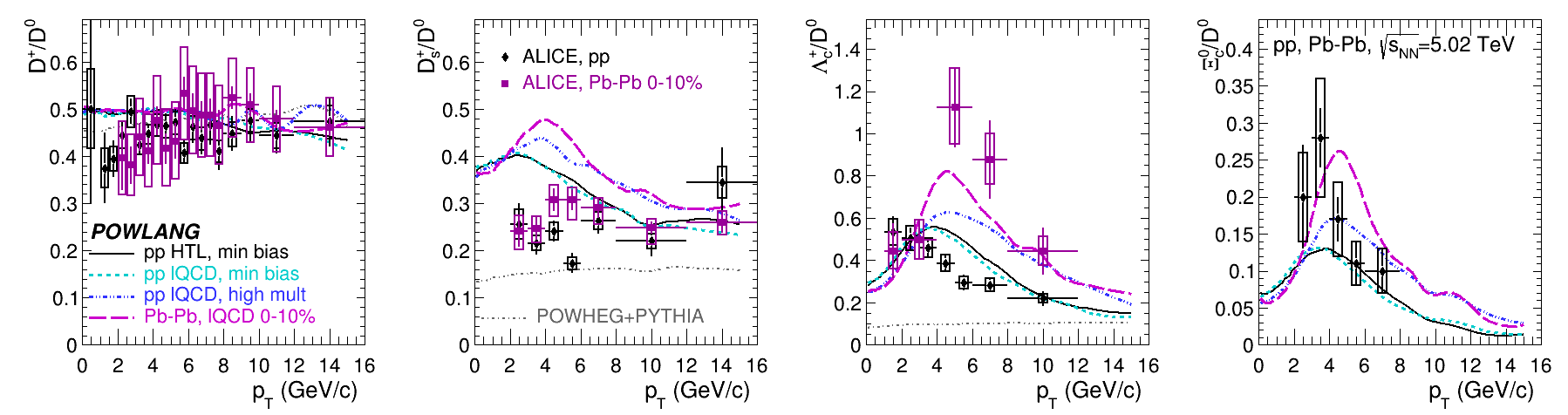}
\caption{Charmed-hadron yield ratios as a function of $p_T$ for different colliding systems at $\sqrt{s_{\rm NN}}\!=\!5.02$ TeV. Predictions including in-medium transport+hadronization in minimum-bias and high-multiplicity pp collisions and in central PbPb collisions (see legend) are compared to ALICE data~\cite{ALICE:2020wfu,ALICE:2021mgk,ALICE:2021rxa,ALICE:2021bib,ALICE:2021psx}. The enhanced baryon-to-meson ratio and the shift of its peak in denser systems is qualitatively well reproduced. Also shown are the pp predictions of POWHEG+PYTHIA standalone, with no medium effects, which undershoots charmed-baryon production.}\label{fig:ratios}
\end{figure*}
An example of the above medium effects found in~\cite{Beraudo:2023nlq} is shown in Fig.~\ref{fig:ratios}, where the ratio of various charmed-hadron species with respect to $D^0$ yields is plotted for different colliding systems: minimum-bias pp collisions, high-multiplicity pp collisions (the 1\% of the events with the highest deposited entropy) and central PbPb collisions. As one can see one gets qualitatively similar results, with an enhanced charmed baryon-to-meson ratio which moves to higher momenta going from minimum-bias pp to central PbPb collisions due to the stronger radial-flow of the fireball. 

\begin{figure*}
\centering
\includegraphics[clip,height=4.5cm]{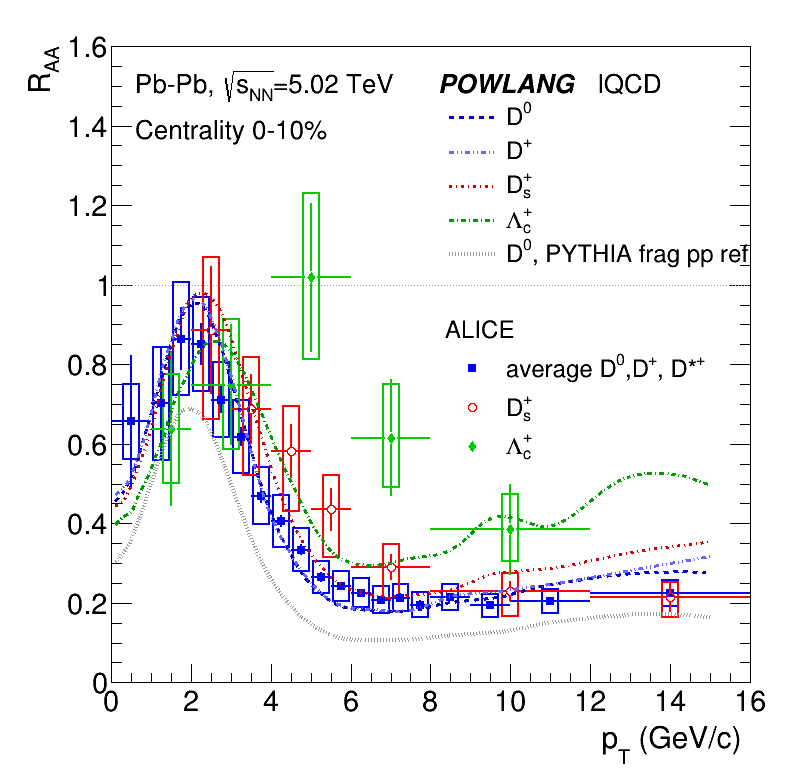}
\hspace{0.3cm}
\includegraphics[clip,height=4.5cm]{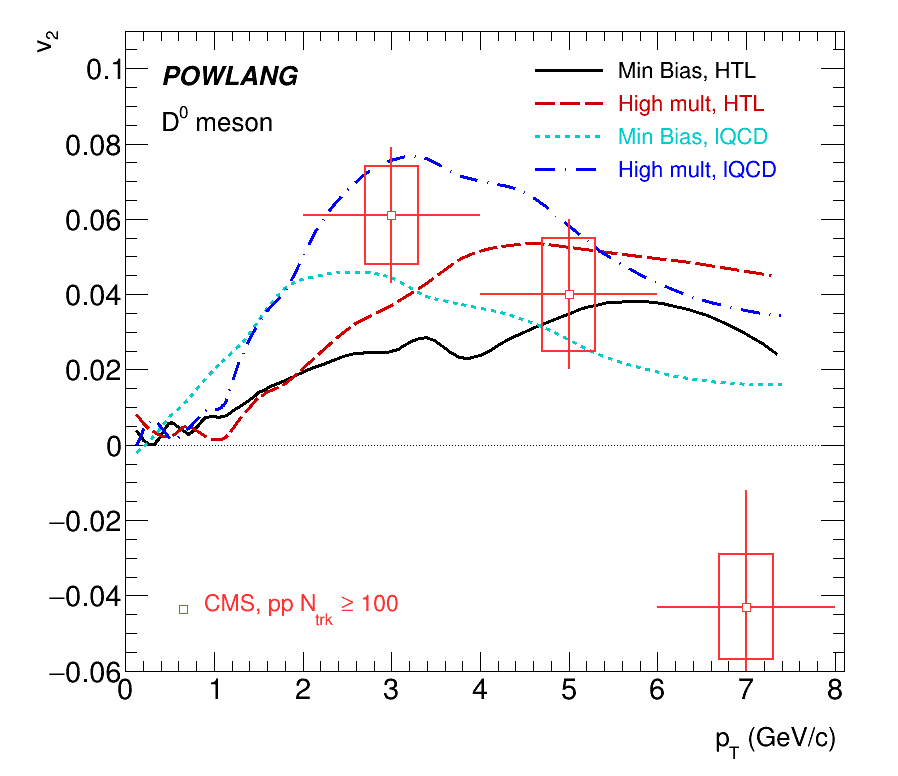}
\caption{Left panel: charmed-hadron nuclear modification factor with (colored curves) and without (grey curve) medium effects in the pp benchmark. Right panel: $D^0$-meson elliptic-flow coefficient in minimum-bias and high-multiplicity pp collisions at $\sqrt{s}\!=\!5.02$ TeV. Theory predictions are compared to CMS results for high-multiplicity pp collisions at $\sqrt{s}\!=\!13$ TeV~\cite{CMS:2020qul}.}\label{fig:pp-RAAv2}
\end{figure*}
This kind of study is also relevant to correctly quantify medium effects in heavy-ion collisions, where the pp benchmark enters in defining the nuclear modification factor $R_{\rm AA}(p_T)\!\propto\!(dN/dp_T)_{\rm AA}/(dN/dp_T)_{\rm pp}$. 
As one can see in the left panel of Fig.~\ref{fig:pp-RAAv2}, the inclusion of medium effects in pp collisions allows one to correctly reproduce the location and magnitude of the radial-flow peak (i.e. the reshuffling of the particle momenta, from low to moderate $p_T$) and to obtain a species dependence of the results with the same qualitative trend of the experimental data. Notice also (see right panel of Fig.~\ref{fig:pp-RAAv2}) that the response to the initial EBE fireball eccentricity leads to a non-vanishing elliptic-flow coefficient $v_2$ of $D^0$ mesons both in minimum-bias and high-multiplicity proton-proton collisions, the response being stronger in the last case due to the longer lifetime of the fireball. An important fraction of the flow is actually acquired at hadronization, due to recombination with nearby thermal particles.

\section{Conclusions}
Even if hadronization is a non-perturbative process associated to one of the most characteristic but hardest features to study of strong interactions -- color confinement -- there is growing evidence that in any hadronic collision the latter occurs via some form of recombination involving partons which initially were not necessarily color-connected, but which eventually are sufficiently close in coordinate and momentum space and with the proper color structure to give rise to low invariant-mass composite color-singlet objects. The advantage of focusing on HF hadrons is that in this case one knows exactly at least one of the parents of the final particle, since charm and bottom quarks can be produced only in the initial hard-scattering or during the parton-shower stage, but not in the final state via excitations of $Q\overline Q$ pairs from the vacuum.

We showed how assuming the formation of a small droplet of QGP also in pp collisions, in which HQ's can undergo rescattering and hadronization, allows one to provide a unified description of a wide set of experimental data concerning HF production, from minimum-bias proton-proton to central nucleus-nucleus collisions. So far most of the studies in the literature were limited to charm. The extension of similar analysis to the bottom sector~\cite{He:2022tod} and also to multi-charmed hadrons~\cite{Minissale:2023bno} is currently addressed by several groups and will contribute to improve our knowledge both on the hadronization mechanism and on the HF transport coefficients in the QGP.

\bibliographystyle{JHEP}
\bibliography{hp23}

\end{document}